\documentclass{llncs}
\usepackage{amsmath}
\usepackage{subfig}
\usepackage{enumerate}

\usepackage{verbatim}
\usepackage{todonotes}
\usepackage{url}

\newcommand{\rdcsim}{RDCSim}

\author{James Stovold \orcidID{0000-0002-0708-2630}} 
\institute{Department of Computer Science \\ Swansea University \\ Swansea, Wales \\ SA1 8EN \\ \url{j.h.stovold@swansea.ac.uk}} 
\title{ \rdcsim{}: a GPU-Accelerated, Interactive Simulator for Reaction--Diffusion Chemistry }

\begin{document}
\maketitle

\begin{abstract} 

This paper presents \rdcsim{}, an interactive simulator for reaction--diffusion chemistry 
(RDC) research, being developed as part of an ongoing project studying how humans interact 
with unconventional computing systems. 

\indent{} While much research into the computational properties of RDC makes use of 
simulations, the development of multiple RDC simulations across different research groups can 
lead to results that are harder to reproduce. By automating the storage of parameter values 
alongside simulation results, \rdcsim{} aims to make reproducing RDC results painless.


\indent{} This paper presents the functionality of \rdcsim{}, and verifies the behaviour of 
the underlying chemical simulation using two seminal examples from the RDC literature: 
logical AND gates and chemical diodes.

\end{abstract}

\section{Introduction}
\label{sec:intro}

Reaction--diffusion chemistry (RDC) is an unconventional computing paradigm that exploits the 
interaction of emergent chemical waves for 
computation~\cite{adamatzky_reactiondiffusion_book,adamatzky_finegrainedcompartmentalised,adamatzky_collisionbased,adamatzky_binarycollisions,agladze_chemicaldiode,delacycostello_towardsconstructing,gorecki_chemicalreactorscount,holley_computationalmodalities,holley_logicalarithmetic_vesicles,stovold_howappropriateare,stovold_simulatingneuronsreaction,stovold_associativememoryreaction,stovold_reactiondiffusionchemistry}. 
This paper presents \rdcsim{}, an interactive simulator for RDC being developed as part of an 
ongoing project looking at how humans interact with unconventional computing systems.


Due to the difficulties associated with producing oscillatory chemical 
reactions~\cite{zhabotinsky_periodicoxidation,zhabotinsky_autowaveprocesses} in a wet lab 
environment, a large number of papers that present results related to RDC make use of 
simulations for their 
experiments~\cite{adamatzky_finegrainedcompartmentalised,adamatzky_collisionbased,delacycostello_towardsconstructing,stovold_howappropriateare,stovold_simulatingneuronsreaction,stovold_associativememoryreaction,stovold_reactiondiffusionchemistry}. 
While there is consensus on the accuracy of the mathematical models compared with the 
underlying chemical reactions, the specific implementations of the models can vary between 
simulations. This increases the risk that results are not consistently reproducible. This 
impact of this risk can be reduced by developing a single, open-source RDC simulator that 
provides complete transparency around parameter values and the underlying chemical model used 
for each simulated result.

\rdcsim{} aims to bring together a single implementation of the most commonly-used RDC models 
and provide functionality to automatically store parameter values alongside results in a 
consistent, structured format. Furthermore, implementing \rdcsim{} on a tablet opens the 
possibility of studying the interaction of humans with unconventional computing substrates.

Finally, \rdcsim{} can be used in research for rapidly designing RDC circuits, and in 
outreach activities to introduce the ideas behind unconventional computing to a wider 
audience.

This paper is organised as follows: section \ref{sec:model} provides the preliminary 
background about RDC, and details the chemical model used by \rdcsim{}; section 
\ref{sec:functionality} describes the functionality of the \rdcsim{} app; section 
\ref{sec:gates} demonstrates the basic correctness of the implemented system by reproducing 
key results from the literature; finally, section \ref{sec:future} discusses future work and 
concludes the paper.

\section{Reaction--Diffusion Chemistry}
\label{sec:model}


 Reaction--diffusion chemistry (RDC) is a chemical reaction that changes state such that 
emergent wavefronts of reagent diffuse across the reactor. The most widely-used models are 
based on the Belousov--Zhabotinsky (BZ)
reaction~\cite{belousov_periodicreaction,zhabotinsky_periodicoxidation}, which may display 
single waves of reaction or an oscillation between two observable states, depending on the 
setup used.

The use of a diffusive chemical reaction as a computational medium has been studied in 
detail, with many image processing operations implemented in the 
substrate~\cite{kuhnert_newopticalmemory,kuhnert_photochemischemanipulation,kuhnert_imageprocessing}. 
The intrinsically parallel nature of RDC makes it an ideal candidate for such operations. 
Other researchers have focussed on the computational power of RDC, looking at maze 
solving~\cite{steinbock_complexlabyrinths}, Voronoi diagram 
construction~\cite{tolmachiev_chemicalprocessor}; chemical 
diodes~\cite{agladze_chemicaldiode}; and involutes~\cite{lazar_involutes}. Recently, much 
attention has been given to the similarities between neuronal behaviour and RDC, including 
spiking neural networks~\cite{stovold_simulatingneuronsreaction}, associative 
memory~\cite{stovold_associativememoryreaction,stovold_reactiondiffusionchemistry}, and 
dynamic neural field theory~\cite{rambidi_molecularnn,stovold_meng}.


In \rdcsim{}, the BZ reaction is simulated using the two-variable Oregonator 
model~\cite{field_oregonator}:
 \begin{equation} \label{eqn:methods:bz}
 \begin{array}{rl}
  \dfrac{\partial u}{\partial t} = & \dfrac{1}{\epsilon}\left[ u - u^{2} - (fv + \phi)\cdot\dfrac{u - q}{u +q}\right] + D_u\nabla^{2}u \\ \\
  \dfrac{\partial v}{\partial t} = & u - v
 \end{array}
\end{equation}
 The default parameter values for the model are given in table~\ref{tab:methods:params}. The 
parameters ($\epsilon$, $f$, $\phi$, $q$, $D_u$) are described in detail 
in~\cite{holley_computationalmodalities}, but (briefly): $\epsilon$ is a scaling factor, $f$ 
is the stoichimetric coefficient (a conservation of mass parameter), $q$ is a propagation 
scaling factor, $\phi$ represents the excitability of the substrate, and $D_u$ is the 
diffusion coefficient for the solution. 

\begin{table}[tbh]
 \centering
 \begin{tabular}{|l|l|}
  \hline
  \textbf\bgroup Parameter \egroup & \textbf\bgroup Value \egroup \\ \hline
  $\epsilon$            & 0.0243                \\ \hline
  $f$                   & 1.4                   \\ \hline
  $\phi_{active}$       & 0.054                 \\ \hline
  $\phi_{passive}$      & 0.0975                \\ \hline
  $q$                   & 0.002                 \\ \hline
  $D_u$                 & 0.45                  \\ \hline
  $\delta t$            & 0.001                 \\ \hline
  $\delta x$            & 0.25                  \\ \hline
 \end{tabular}
 \vspace{10pt}
 \caption[Parameter Values]{Parameter values for excitable Belousov--Zhabotinsky reaction 
simulated through the Oregonator model.}
 \label{tab:methods:params}
\end{table}

The model in eqn.~\ref{eqn:methods:bz} is numerically integrated using an explicit forward 
Euler integration, with timestep $\delta t=0.001$ and five-node discrete Laplace operator 
with grid spacing $\delta x=0.25$. Catalytic stimulation is achieved by setting the 
excitatory concentration ($u$) of stimulated pixels to 1.0.


The implemented model is accelerated using a GPU, making the simulation sufficiently 
responsive for user interaction. Due to this acceleration, the user is able to vary the 
observable speed of the simulation according to preference or experimental requirement. This 
is achieved by varying how often the reaction is rendered to screen.

\section{Functionality of the Simulator}
\label{sec:functionality}

\rdcsim{} is built as an iPad app, making use of the built in GPU for accelerating the 
chemical model. All testing has been performed on the 2017 10.5in iPad Pro running iOS 12.0. 
This tablet runs an Apple A10X ARM CPU, with 6 cores running at 2.38GHz, 4GB RAM, and a 
12-core GPU. By building \rdcsim{} as an app, the touchscreen capabilities of a tablet make 
designing and running RDC experiments substantially easier. The user is able to stimulate the 
reaction with a catalyst using their finger or stylus, and are able to draw circuits in the 
same way.

\begin{figure}[h!]
 \centering 
 \includegraphics[width=0.8\linewidth]{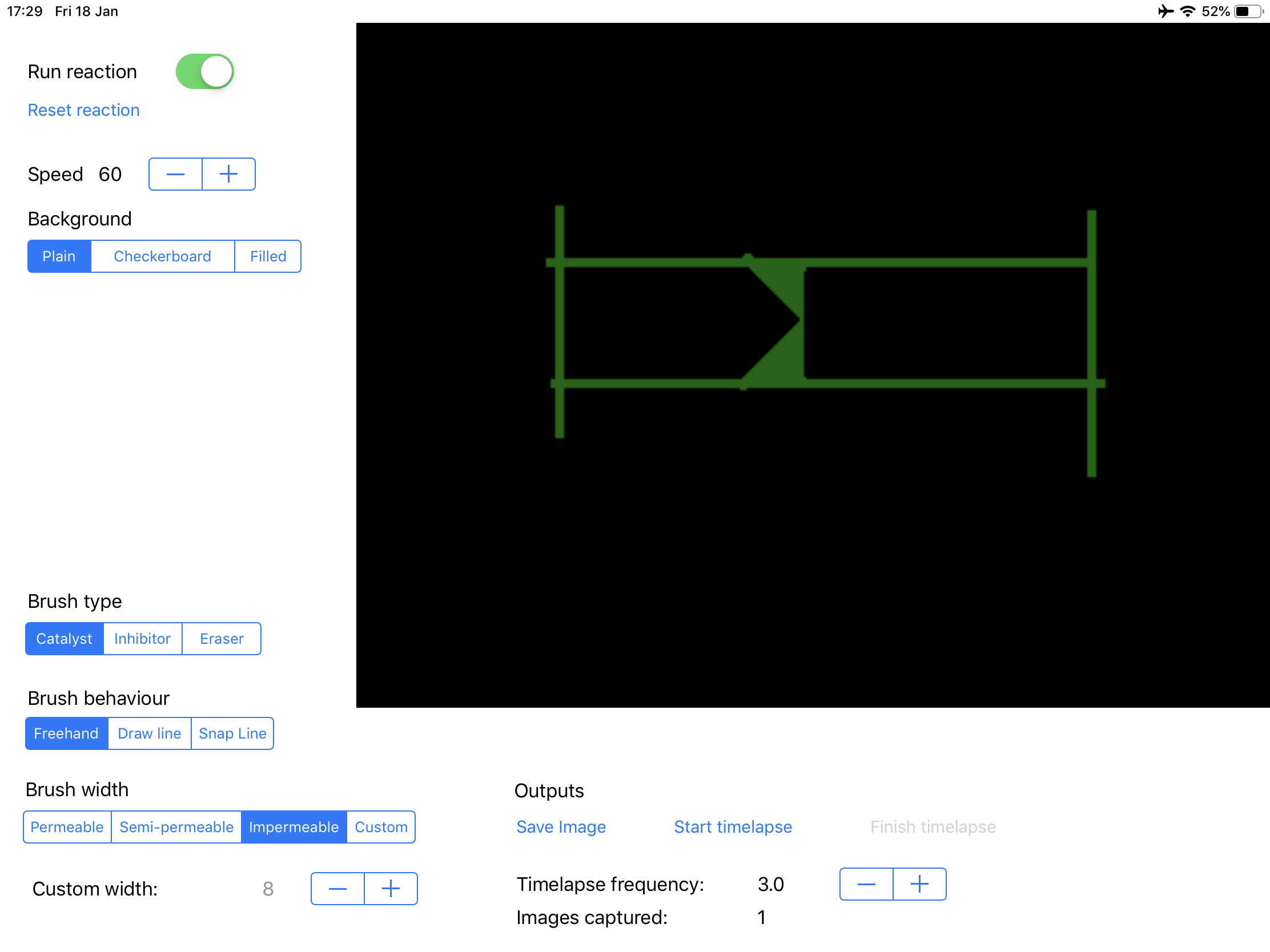}
 \caption[]{User interface of \rdcsim{}. The main reaction takes place in a 400x300 pixel 
simulation, scaled to 800x600 pixels for display to the user.}
 \label{fig:interface}
\end{figure}

The \rdcsim{} app, the interface for which is shown in fig.~\ref{fig:interface}, has a range 
of functionality implemented with much more planned for development in the near future. The 
order with which this future functionality is implemented will depend on the requirements 
when running human experiments and on specific requests from other researchers using the app.

At present, the system allows for freehand and line drawing of inhibitive substrate and 
catalytic stimulation. Furthermore, to aid in the development of logical circuits, a `snap 
line' option has been introduced that draws a line restricted to multiples of 15-degree 
angles only.

The width of the lines drawn can be customised by the user, with presets implemented that 
permit different degrees of wave permeability. `Permeable' lines only slow the propagating 
waves; `semi-permeable' lines allow waves to pass that approach perpendicular to the barrier, 
but not parallel, making them ideal for logical AND gates (see sect.~\ref{sec:gates:and}); 
`impermeable' line stop all waves. In the presets for \rdcsim{}, these are defined as 2, 3, 
and 8 pixels wide for permeable, semi-permeable, and impermeable lines, respectively.

Furthermore, presets for the background have been implemented so that the user can choose to 
work by removing substrate from a filled background or by adding it to an empty background. 
There are plans to implement saving and loading of user-defined backgrounds in the near 
future. Fig.~\ref{fig:interface} shows the user interface of the system, with an empty 
background (black) and a diode circuit being tested.

As described in section~\ref{sec:model} the chemical model currently implemented in \rdcsim{} 
is the two-variable Oregonator model~\cite{field_oregonator}, based on the well-known 
Belousov--Zhabotinsky (BZ) 
reaction~\cite{zhabotinsky_periodicoxidation,zhabotinsky_autowaveprocesses}. At present, the 
variables for this model are fixed but work is ongoing to make these available to change by 
the user, along with providing presets for different observed behaviours (such as 
sub-excitable wave fragments~\cite{adamatzky_binarycollisions}).

Finally, one of the key objectives for this project is to increase the reproducibility of 
results in RDC research, so the next major release of \rdcsim{} will include database 
integration. This will provide a way to consistently save results and parameter values 
alongside one another. Moreover, by using a structured format for parameters and results, 
these can be re-loaded into \rdcsim{} for verification at a later date.

\subsection{Outputs from the Simulator}
\label{sec:outputs}


The simulated reaction is rendered to screen using the following colour-mapping: 

\begin{itemize} 
 \item[] Red pixels: concentration of $u$
 \item[] Green pixels: inhibitive substrate
 \item[] Blue pixels: concentration of $v$
\end{itemize}

\begin{figure}[h!]
 \centering
 \includegraphics[width=0.5\linewidth]{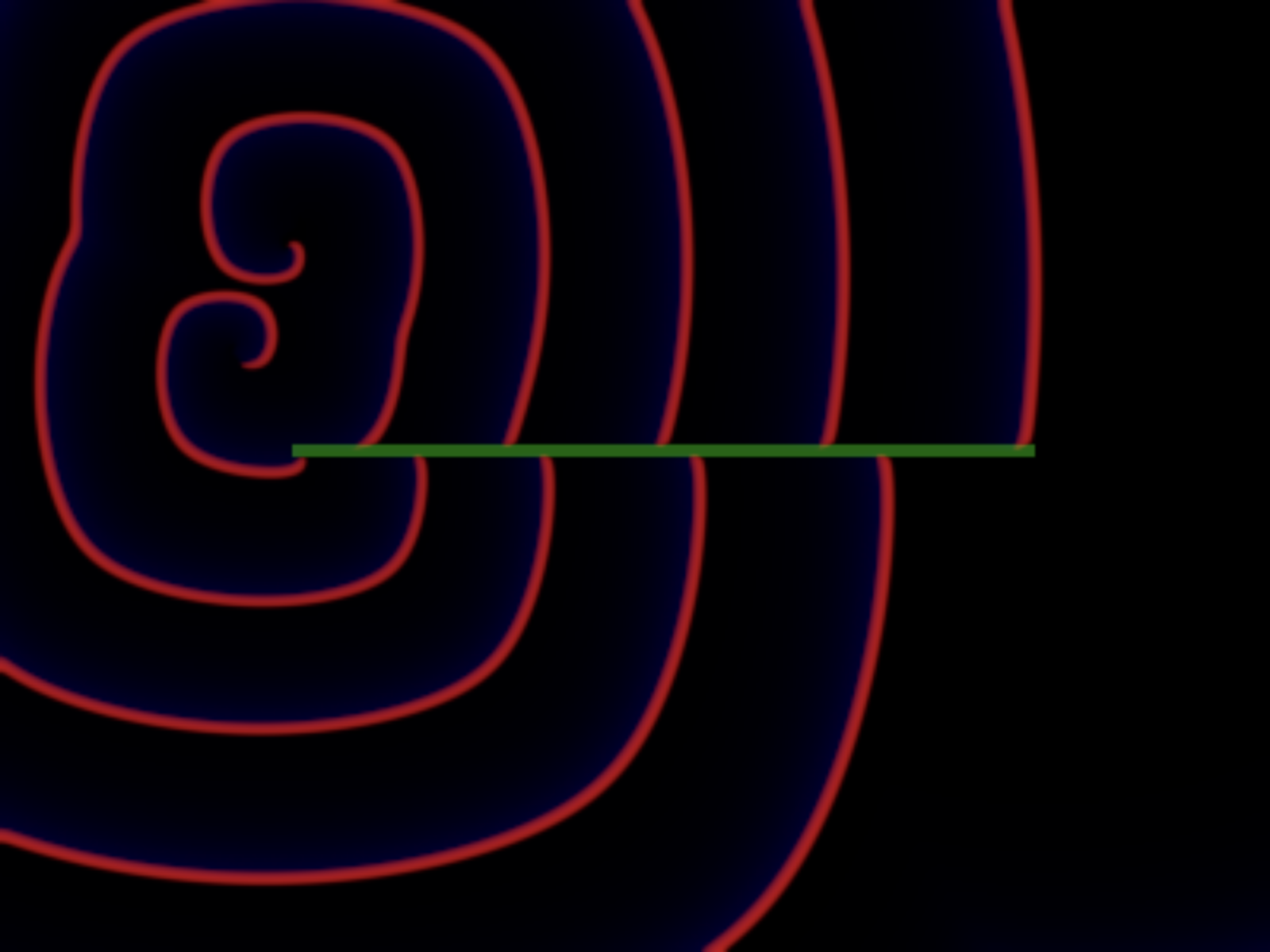}
 \caption[]{Spiral pattern producing a series of waves, half of which are delayed by the 
inhibitive substrate (green).}
 \label{fig:output:spiral}
\end{figure}

Fig.~\ref{fig:output:spiral} shows an image rendered by the app, displaying alternating waves 
above and below the inhibitor, produced by the dual spiral patterns in the top-left.
\vspace{-1.5em}

\begin{figure}[h!]
 \centering 
\subfloat[Snapshot output \label{fig:sample:snapshot} ]{ \centering
  \includegraphics[width=0.4\linewidth]{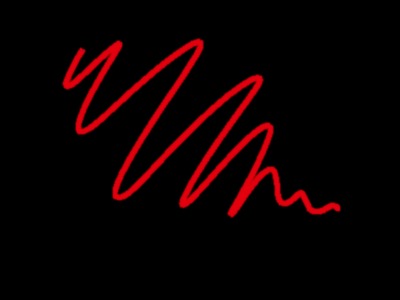}
}
\subfloat[Timelapse output \label{fig:sample:timelapse} ]{ \centering
  \includegraphics[width=0.4\linewidth]{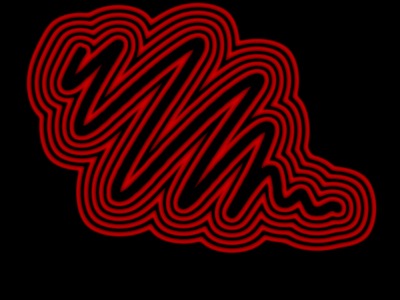}
}
 \caption[]{The two main forms of visual output from \rdcsim{}. (a) shows an early-stage 
hand-drawn stimulus, (b) shows the timelapse of how the hand-drawn stimulus propagates across 
the reactor. }
 \label{fig:sample_output}
\end{figure}

\vspace{-1em}
There are currently two ways to output results from the simulator. The first saves a snapshot 
of the current state of the reaction to an image on the tablet. The second records a 
timelapse of the reaction at a rate set by the user. This timelapse can then be saved as an 
image to the tablet. In order to improve the clarity of timelapse images, only the 
concentration of $u$ is recorded over time, and overlaid on top of the static inhibitive 
substrate. An example snapshot and timelapse image are provided in 
figs.~\ref{fig:sample:snapshot} and \ref{fig:sample:timelapse}.

\vspace{-1em}

\newpage{}

\section{Basic Behaviour Verification}
\label{sec:gates}

To demonstrate that the system is working correctly, this section reproduces two seminal RDC 
results from the literature: the logical AND gate~\cite{gorecki_chemicalreactorscount} and 
chemical diode~\cite{agladze_chemicaldiode}. All circuits and catalytic stimulation used in 
this section have been drawn by hand using the tablet interface, and results are presented 
using the standard outputs from the simulator.

\subsection{Logical AND Gate}
\label{sec:gates:and}

\begin{figure}[h!]
 \centering
 \includegraphics[width=\linewidth]{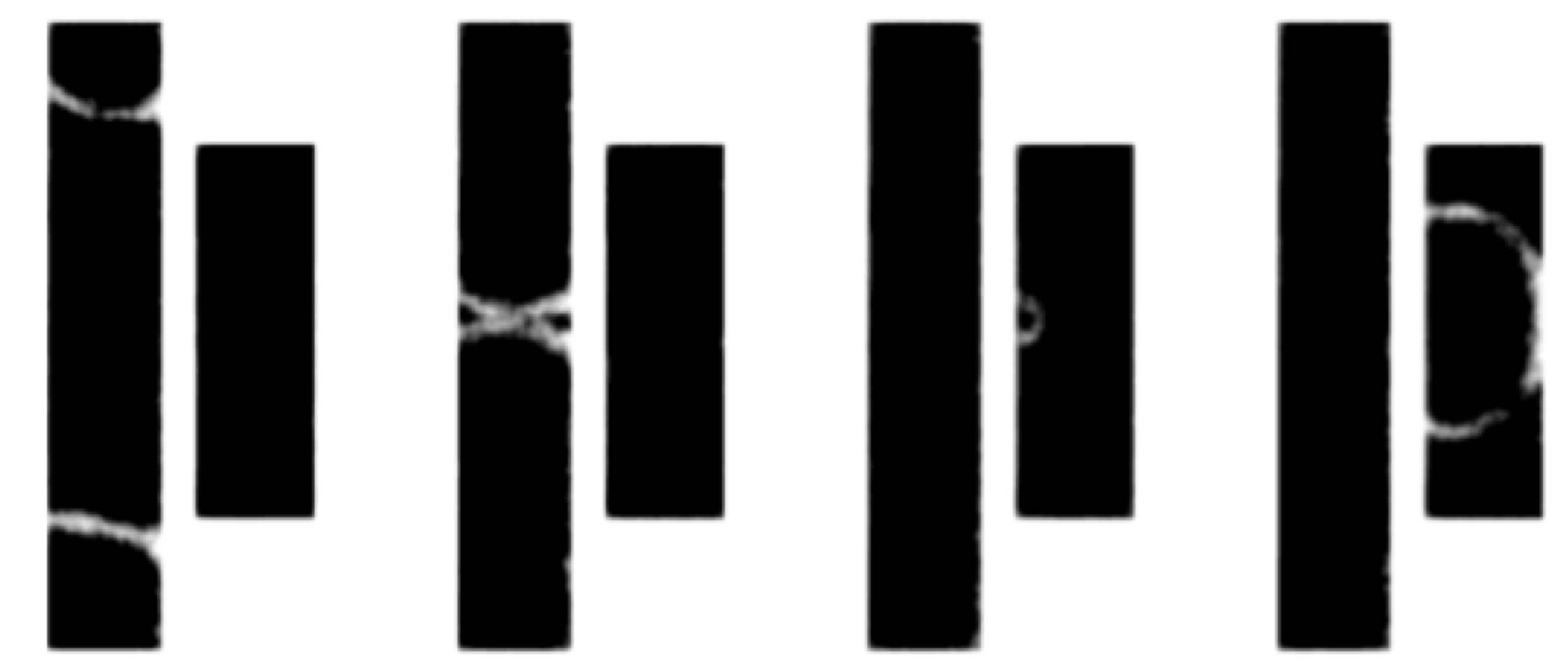}
 \caption[]{Four stages of Gorecki et al.'s coincidence detector in chemical medium. Time 
moves from left to right, showing the interaction of two chemical waves and the resulting 
wave propagating to the output. Image reproduced from~\cite{gorecki_chemicalreactorscount}.  
}
 \label{fig:gates:original}
\end{figure}

Due to the lack of a clock signal in RDC, a logical AND gate is equivalent to a coincidence 
detector~\cite{gorecki_chemicalreactorscount}. By arranging two active regions separated by a 
passive region (as shown in fig.~\ref{fig:gates:original}), the wave can only propagate 
across the passive region if two waves collide opposite the output region.

\begin{figure}[h!]
 \centering
 \subfloat[ AND gate with inputs 01 produces a 0 as output. \label{fig:gates:and:01} ] {\centering
 \includegraphics[width=0.45\linewidth]{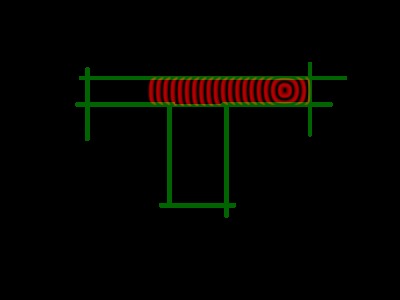}
 } ~
 \subfloat[ AND gate with inputs 11 produces a 1 as output. \label{fig:gates:and:11} ] {\centering
 \includegraphics[width=0.45\linewidth]{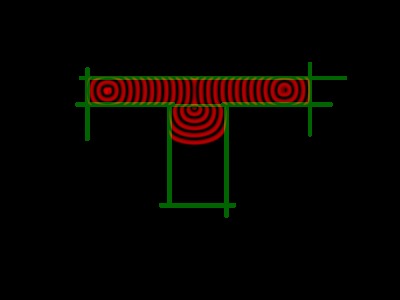}
 } 
 \caption[]{Timelapse images of \rdcsim{}-implemented AND gates. The successful 
implementation of a logical AND gate demonstrates that the simulation of an RDC system is 
working as expected.}
 \label{fig:gates:and}
\end{figure}

Fig.~\ref{fig:gates:and:01} shows a timelapse image of a single input entering the 
implemented gate, demonstrating that it is not able to propagate to the bottom of the 
circuit, where the output is measured. When both inputs are present, however, as in 
fig.~\ref{fig:gates:and:11}, the interaction between the two waves is sufficient to stimulate 
the output region and form an output wave. As such, the conjunctive logic is implemented and 
successfully demonstrated.

\subsection{Chemical Diode}
\label{sec:gates:diode}

\begin{figure}[h!]
 \centering
 \includegraphics[width=\linewidth]{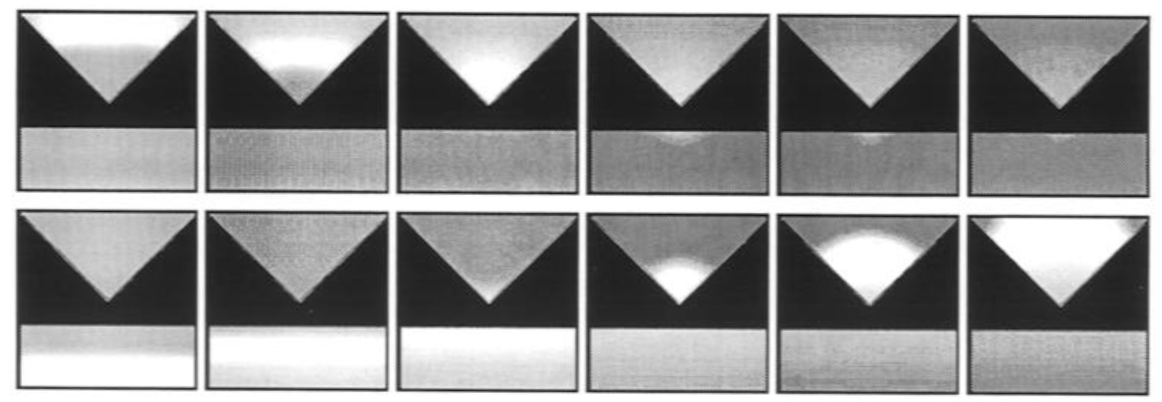}
 \caption[]{ Agladze et al.'s original chemical diode~\cite{agladze_chemicaldiode}. The top 
row shows the wave---approaching from the top---unable to propagate from cathode to anode. 
The bottom row shows the wave---approaching from below---able to propagate from anode to 
cathode. Time progresses from left to right. Image reproduced 
from~\cite{agladze_chemicaldiode}.}
 \label{fig:gates:diode:original}
\end{figure}

The chemical diode~\cite{agladze_chemicaldiode} is an essential component for any medium- or 
large-sized RDC circuit. It prevents waves from propagating in an unintended direction 
through the circuit. The basic chemical diode is a very simple design (see 
fig.~\ref{fig:gates:diode:original}), consisting of a gap in active substrate with a pointed 
interface on the cathode side of the diode and a flat interface on the anode side of the 
diode.\footnote{assuming the chemical waves follow conventional current, although arguably 
the emergent waves in an RDC system are more similar to voltage waves, which would reverse 
the description of anode/cathode arrangement presented here, but not the behaviour}

\begin{figure}[h!]
 \centering
 \subfloat[ Chemical diode with wave propagating from anode to cathode. \label{fig:gates:diode:right} ] {\centering
  \includegraphics[width=0.45\linewidth]{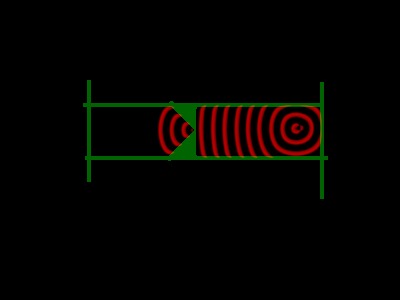}
 } ~
 \subfloat[ Chemical diode with wave failing to propagate from cathode to anode. 
\label{fig:gates:diode:left} ] {\centering
  \includegraphics[width=0.45\linewidth]{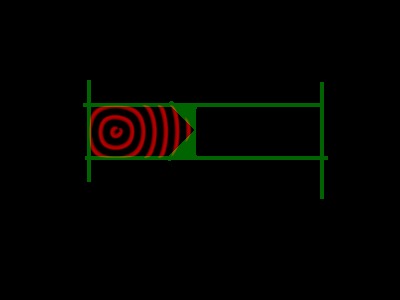}
 }
 \caption[]{Timelapse images of \rdcsim{}-implemented diodes. The successful implementation 
of a logical diode demonstrates that the simulation of an RDC system is working as expected.}
 \label{fig:gates:diode}
\end{figure}

This design for a diode is implemented in \rdcsim{} and allows chemical waves to propagate 
from the anode to the cathode, but not from the cathode to the anode---as shown in 
figs.~\ref{fig:gates:diode:right} and \ref{fig:gates:diode:left}.

\section{Future Work}
\label{sec:future}

While the simulation is currently sufficient to run simple experiments, there are a 
substantial number of improvements planned for the near future. The first improvement is 
already under development, to allow users to alter the chemical model parameters and to 
integrate the simulation with a database to store the parameter values alongside simulation 
outputs. This will help to improve consistent reproducibility of results, one of the primary 
aims for developing \rdcsim{}.

Furthermore, the ability for users to save and load presets, and to save and load circuits / 
backgrounds is a high priority, again helping to improve reproducibility across experiments, 
but also to improve the speed with which researchers can develop RDC solutions.

Finally, the long-term plans for \rdcsim{} include development of multiple chemical models, 
vesicle models, and the ability to record inputs in real-time. These will help to expand the 
user base for the simulator, but are not essential for current research plans.

The primary application for this \rdcsim{} is for the study of human--RDC interaction. By 
studying how lay humans naturally interact with a chemical substrate, novel forms of 
human--computer interaction could be formed which can help inform current practice. The use 
of tangible chemical interfaces are already being 
investigated~\cite{sahoo_tangibledropsvisio,tokuda_programmableliquidmatter,tokuda_programmableliquidmatter2}, 
and combining the computational power of diffusive computing with the interfaces already 
under development offers a route towards radically-different computing systems in future.

\section{Summary}
\label{sec:summary}

This paper has presented the ongoing development work of \rdcsim{}, a GPU-accelerated, 
interactive simulator for reaction--diffusion chemistry research. The simulator is verified 
using two seminal examples from the RDC literature: logical AND gates and chemical diodes.

There are two main aims for \rdcsim{}. The first is to develop a simulator that enables 
effortlessly-reproducible RDC research. The second is to use the app to study how humans 
interact with unconventional computing systems. The work in this paper represents the first 
step towards these aims, providing a system that can be used for basic human interaction 
studies.

Finally, the author hopes that other RDC researchers will make use of, and contribute to, the 
development of the simulator\footnote{the code for \rdcsim{} is open-source and available at 
\\ \url{https://github.com/jstovold/RDCSimulator}}, and that through various outreach events 
it can be used to encourage new interest in unconventional computing methods.

\begin{thebibliography}{10}
\providecommand{\url}[1]{\texttt{#1}}
\providecommand{\urlprefix}{URL }
\providecommand{\doi}[1]{https://doi.org/#1}

\bibitem{adamatzky_reactiondiffusion_book}
Adamatzky, A., De~Lacy~Costello, B., Asai, T.: Reaction--Diffusion Computers.
  Elsevier Science (2005)

\bibitem{adamatzky_finegrainedcompartmentalised}
{Adamatzky}, A., {Holley}, J., {Bull}, L., {de Lacy Costello}, B.: {On
  computing in fine-grained compartmentalised Belousov--Zhabotinsky medium}.
  Chaos Solitons \& Fractals  \textbf{44},  779--790 (Oct 2011)

\bibitem{adamatzky_collisionbased}
Adamatzky, A.: Collision-based computing in {Belousov}--{Zhabotinsky} medium.
  Chaos, Solitons \& Fractals  \textbf{21}(5),  1259--1264 (2004)

\bibitem{adamatzky_binarycollisions}
Adamatzky, A., {De Lacy Costello}, B.: Binary collisions between wave-fragments
  in a sub-excitable {Belousov}--{Zhabotinsky} medium. Chaos, Solitons \&
  Fractals  \textbf{34}(2),  307--315 (2007)

\bibitem{agladze_chemicaldiode}
Agladze, K., Aliev, R.R., Yamaguchi, T., Yoshikawa, K.: Chemical diode. J.
  Phys. Chem  \textbf{100},  13895--13897 (1996)

\bibitem{belousov_periodicreaction}
Belousov, B.P.: A periodic reaction and its mechanism. Med. Publ., Moscow
  (1959)

\bibitem{delacycostello_towardsconstructing}
Costello, B.D.L., Adamatzky, A., Jahan, I., Zhang, L.: Towards constructing
  one-bit binary adder in excitable chemical medium. Chemical Physics
  \textbf{381}(1),  88--99 (2011)

\bibitem{field_oregonator}
Field, R.J., Janz, R.D., Vanecek, D.J.: Composite double oscillation in a
  modified version of the {Oregonator} model of the {Belousov--Zhabotinsky}
  reaction. The Journal of Chemical Physics  \textbf{73}(7),  3132--3138 (1980)

\bibitem{gorecki_chemicalreactorscount}
Gorecki, J., Yoshikawa, K., Igarashi, Y.: On chemical reactors that can count.
  The Journal of Physical Chemistry A  \textbf{107}(10),  1664--1669 (2003)

\bibitem{holley_computationalmodalities}
{Holley}, J., {Adamatzky}, A., {Bull}, L., {De Lacy Costello}, B., {Jahan}, I.:
  Computational modalities of {Belousov--Zhabotinsky} encapsulated vesicles.
  ArXiv e-prints  (Sep 2010)

\bibitem{holley_logicalarithmetic_vesicles}
Holley, J., Jahan, I., De~Lacy~Costello, B., Bull, L., Adamatzky, A.: Logical
  and arithmetic circuits in {Belousov--Zhabotinsky} encapsulated disks. Phys.
  Rev. E  \textbf{84},  056110 (Nov 2011)

\bibitem{kuhnert_newopticalmemory}
Kuhnert, L.: A new optical photochemical memory device in a light-sensitive
  chemical active medium. Nature  \textbf{319}(6052),  393--394 (Jan 1986)

\bibitem{kuhnert_photochemischemanipulation}
Kuhnert, L.: Photochemische manipulation von chemischen wellen (in {German}).
  Naturwissenschaften  \textbf{73},  96--97 (1986)

\bibitem{kuhnert_imageprocessing}
Kuhnert, L., Agladze, K.I., Krinsky, V.I.: Image processing using
  light-sensitive chemical waves. Nature  \textbf{337}(6204),  244--247 (Jan
  1989)

\bibitem{lazar_involutes}
L\'{a}z\'{a}r, A., Noszticzius, Z., Farkas, H., F\"{o}rsterling, H.D.:
  Involutes: the geometry of chemical waves rotating in annular membranes.
  Chaos  \textbf{5}(2),  443--447 (1995)

\bibitem{rambidi_molecularnn}
Rambidi, N., Maximychev, A., Usatov, A.: Molecular neural network devices based
  on non-linear dynamic media. Biosystems  \textbf{33}(2),  125--137 (1994)

\bibitem{sahoo_tangibledropsvisio}
Sahoo, D.R., Neate, T., Tokuda, Y., Pearson, J., Robinson, S., Subramanian, S.,
  Jones, M.: Tangible drops: A visio-tactile display using actuated
  liquid-metal droplets. In: Proceedings of the 2018 CHI Conference on Human
  Factors in Computing Systems. pp. 177:1--177:14. CHI '18, ACM (2018)

\bibitem{steinbock_complexlabyrinths}
Steinbock, O., T{\'o}th, {\'A}., Showalter, K.: Navigating complex labyrinths:
  Optimal paths from chemical waves. Science  \textbf{267}(5199),  868--871
  (1995)

\bibitem{stovold_meng}
Stovold, J.: Extending the Computational Application of Reaction--Diffusion
  Chemistry by Modelling Artificial Neural Networks. Master's thesis,
  University of York (2012)

\bibitem{stovold_howappropriateare}
Stovold, J., O'Keefe, S.: How appropriate are logic gates in
  reaction--diffusion chemistry? In: Bandur, V. (ed.) Proc. 5th York Doctoral
  Symposium on Computer Science. vol.~5, pp. 17--26. University of York (2012)

\bibitem{stovold_simulatingneuronsreaction}
Stovold, J., O'Keefe, S.: Simulating neurons in reaction-diffusion chemistry.
  In: Lones, M.A., Smith, S.L., Teichmann, S., Naef, F., Walker, J.A., Trefzer,
  M.A. (eds.) Proc. IPCAT 2012, LNCS 7223. pp. 143--149 (2012)

\bibitem{stovold_associativememoryreaction}
Stovold, J., O'Keefe, S.: Associative memory in reaction--diffusion chemistry.
  In: Adamatzky, A. (ed.) Advances in Unconventional Computing: Volume 2:
  Prototypes, Models and Algorithms, pp. 141--166. Springer International
  Publishing (2017)

\bibitem{stovold_reactiondiffusionchemistry}
Stovold, J., O'Keefe, S.: Reaction--diffusion chemistry implementation of
  associative memory neural network. International Journal of Parallel,
  Emergent and Distributed Systems  \textbf{32}(1),  74--94 (2017)

\bibitem{tokuda_programmableliquidmatter2}
Tokuda, Y., Moya, J.L.B., Memoli, G., Neate, T., Sahoo, D.R., Robinson, S.,
  Pearson, J., Jones, M., Subramanian, S.: Programmable liquid matter: 2d shape
  deformation of highly conductive liquid metals in a dynamic electric field.
  In: Proceedings of the 2017 ACM International Conference on Interactive
  Surfaces and Spaces. pp. 142--150. ISS '17, ACM (2017)

\bibitem{tokuda_programmableliquidmatter}
Tokuda, Y., Moya, J.L.B., Memoli, G., Neate, T., Sahoo, D.R., Robinson, S.,
  Pearson, J., Jones, M., Subramanian, S.: Programmable liquid matter: 2d shape
  drawing of liquid metals by dynamic electric field. In: Proceedings of the
  2017 ACM International Conference on Interactive Surfaces and Spaces. pp.
  454--457. ISS '17, ACM (2017)

\bibitem{tolmachiev_chemicalprocessor}
Tolmachiev, D., Adamatzky, A.: Chemical processor for computation of {Voronoi}
  diagram. Advanced Materials for Optics and Electronics  \textbf{6}(4),
  191--196 (1996)

\bibitem{zhabotinsky_periodicoxidation}
Zhabotinsky, A.M.: Periodic course of the oxidation of malonic acid in a
  solution (studies on the kinetics of {Belousov}'s reaction). Biofizika
  \textbf{9} (1964)

\bibitem{zhabotinsky_autowaveprocesses}
Zhabotinsky, A., Zaikin, A.: Autowave processes in a distributed chemical
  system. Journal of Theoretical Biology  \textbf{40}(1),  45--61 (1973)

\end{thebibliography}

\end{document}